\documentclass[conference]{IEEEtran}
\IEEEoverridecommandlockouts
\usepackage{amsmath,amssymb,amsfonts}
\usepackage{graphicx}
\graphicspath{{assets/}{assets/unlimited/}}
\usepackage{textcomp}
\usepackage[hyphens]{url}
\usepackage{fancyhdr}
\usepackage{hyperref}
\usepackage{multirow}
\usepackage{xcolor,colortbl}
\usepackage{booktabs}
\usepackage{bookmark}
\usepackage{cleveref}
\usepackage[final]{microtype}
\usepackage[a4paper, total={184mm,239mm}]{geometry}
\def\BibTeX{{\rm B\kern-.05em{\sc i\kern-.025em b}\kern-.08em
    T\kern-.1667em\lower.7ex\hbox{E}\kern-.125emX}}

\newcommand{\minisection}[1]{\vspace{.06in}\noindent{\textbf{#1}}.}

\usepackage{algorithm}
\usepackage{algpseudocode}

\begin{document}

\emergencystretch=3em

\title{AnchorTP: Resilient LLM Inference with State-Preserving Elastic Tensor Parallelism}

\author{Wendong Xu\textsuperscript{1,*}, Chujie Chen\textsuperscript{2}, He Xiao\textsuperscript{1}, Kuan Li\textsuperscript{3}, Jing Xiong\textsuperscript{1}, Chen Zhang\textsuperscript{1}, \\Wenyong Zhou\textsuperscript{1}, Chaofan Tao\textsuperscript{1}, Yang Bai\textsuperscript{4}, Bei Yu\textsuperscript{4}, Ngai Wong\textsuperscript{1}\\

\textsuperscript{1}Department of Electrical and Electronic Engineering, The University of Hong Kong, Hong Kong \\
\textsuperscript{2}Institute of Computing Technology, Chinese Academy of Sciences, Beijing, China \\
\textsuperscript{3}Department of Computer Science and Engineering, Hong Kong University of Science and Technology, Hong Kong \\
\textsuperscript{4}Department of Computer Science and Engineering, Chinese University of Hong Kong, Hong Kong \\
\textsuperscript{*}Corresponding author. Email: wdxu@connect.hku.hk
}

\maketitle

\begin{abstract}
Large Language Model (LLM) inference services demand exceptionally high availability and low latency, yet multi-GPU Tensor Parallelism (TP) makes them vulnerable to single-GPU failures. We present AnchorTP, a state-preserving elastic TP framework for fast recovery. It (i) enables Elastic Tensor Parallelism (ETP) with unequal-width partitioning over any number of GPUs and compatibility with Mixture-of-Experts (MoE), and (ii) preserves model parameters and KV caches in GPU memory via a daemon decoupled from the inference process. To minimize downtime, we propose a bandwidth-aware planner based on a Continuous Minimal Migration (CMM) algorithm that minimizes reload bytes under a byte-cost dominance assumption, and an execution scheduler that pipelines P2P transfers with reloads.
These components jointly restore service quickly with minimal data movement and without changing service interfaces. In typical failure scenarios, AnchorTP reduces Time to First Success (TFS) by up to 11$\times$ and Time to Peak (TTP) by up to 59\% versus restart-and-reload.
\end{abstract}
\section{Introduction}
\label{sec:introduction}

Online large language models (LLMs) serving systems have become critical infrastructure~\cite{zheng2024sglang,kwon2023efficient}. To meet compute and memory capacity demands, inference commonly adopts tensor parallelism (TP) across multiple GPUs~\cite{wang2022tesseract,brakel2024model,wu2024loongserve}. The tight coupling in TP makes systems highly sensitive to single-GPU failures or link degradation~\cite{arfeen2025nonuniform,cui2025characterizing}: once a communication group breaks, service is interrupted and the time to first success (TFS) can reach tens of minutes~\cite{blagoev2025all,xia2025mnemosyne}. This work aims to reduce TFS (time from fault to first successful token; often called TTFT) to seconds while preserving throughput and latency, and to shorten TTP (time from first success to stabilized peak throughput).

However, existing TP implementations hardwire divisibility constraints on key dimensions, fix per-GPU tensor shapes, and assume a static scale of collective communications. When topology changes, shapes and collectives globally mismatch, leading to high reconfiguration costs.

\begin{figure}[t]
    \centering
    \includegraphics[width=0.9\linewidth]{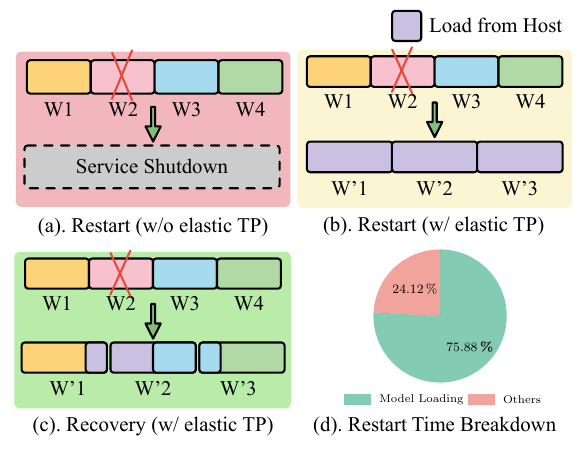}
    \vspace{-1.5em}
    \caption{Recovery strategies when one GPU fails in a four-GPU deployment. (a) Without elastic TP, service cannot resume. (b) With elastic TP but no state preservation, service restarts with three GPUs but fully reloads parameters from host. (c) With state-preserving elastic TP, parameters/KVs on surviving GPUs are reused via planned P2P transfers with minimal reload. (d) Time breakdown for a typical restart-and-reload on Qwen3-14B~\cite{qwen3technicalreport}; host-to-GPU reload dominates.}
    \label{fig:recovery_illustration}
    \vspace{-1.5em}
\end{figure}

To achieve fast recovery, the industry typically employs two main approaches: restart-and-reload, which simplifies the process but yields long TFS/TTP~\cite{strati2024d,wu2023transom,xie2025realm}; and static redundancy, which relies on standby nodes or replicas, sacrificing flexibility and cost efficiency. Practice reveals three challenges~\cite{cheng2025scalable,xu2025cloud,zhang2024edgeshard}: (i) topology-aware migration costs under bandwidth tiering and limited reachability, (ii) compute–communication rebalancing after TP-scale changes, and (iii) runtime state persistence where KVs, memory pools, and communication groups are tightly bound to the initial topology. These constraints shrink the feasible design space and motivate maximizing in-place reuse while minimizing necessary migrations (see \Cref{fig:recovery_illustration}).

As illustrated in \Cref{fig:recovery_illustration}(d), restart-and-reload is bottlenecked by host-to-GPU model reload time. We observe that significant memory redundancy in production deployments creates an opportunity for low-cost state-preserving recovery.

The widespread adoption of Mixture-of-Experts (MoE) architectures~\cite{jacobs1991adaptive,masoudnia2014mixture,liu2024deepseek} has made sparse activation and expert routing the norm, further tightening the coupling between state and communication and raising the requirements for KV-state consistency and topology reconfiguration. Expert-parallel load balancing (EPLB)~\cite{guo2025deepseek,wisdom2024load} primarily focuses on optimizing routing popularity and expert utilization, which is essentially dynamic load balancing.
EPLB strategies rely on the recency and temporal coherence of request traffic; once a service restart\-and\-reload occurs, this coherence is disrupted, which can bias estimates of traffic distribution and hot experts~\cite{zeng2025efficientmoe}. As a result, larger scheduling and routing adjustments may be triggered, which in the short term exacerbate expert load imbalance and degrade performance.

To address these challenges, we propose AnchorTP, a disaster-resilient elastic inference framework. AnchorTP decouples state from control: the state plane persistently holds GPU memory ownership of model parameters and KVs, while the control plane drives elastic TP reconfiguration and minimal-migration recovery after failures. Without changing the service interface, we convert costly reloads into a small number of bandwidth-aware migrations (\Cref{fig:recovery_illustration}(b)--(c)). We focus on multi-GPU inference under device-offline and link-degradation scenarios, targeting low TFS and reduced TTP by maximizing in-place reuse and minimizing necessary migrations. While our evaluation concentrates on single-node deployments, the core principles extend to multi-node environments where the primary difference lies in communication overhead rather than algorithmic complexity.

We summarize our contributions as follows:
\begin{itemize}
    \item We decouple a state plane and a control plane. The state plane anchors GPU memory for parameters and KVs via a daemon and IPC handles, enabling in-place fast recovery.
    \item We implement Elastic Tensor Parallelism (ETP) with unequal-width sharding for attention and linear layers. ETP breaks divisibility constraints, supports arbitrary TP-scale reconfiguration, and remains compatible with MoE.
    \item We design a Continuous Minimal Migration (CMM) planner with bandwidth-aware execution. It minimizes reload bytes by interval mapping under the cost model and overlaps P2P with reloads to reduce wall-clock time.
\end{itemize}

\section{Preliminaries}
\label{sec:preliminaries}

\subsection{Parallel Topologies}
Distributed inference for LLMs typically employs a hybrid parallel strategy combining Tensor Parallelism (TP) and Expert Parallelism (EP) in MoE models~\cite{singh2023hybrid,zhu2025megascale}. TP addresses single-card memory bottlenecks by partitioning linear layer computations across multiple GPUs, while EP distributes different experts across devices to achieve sparse activation of computation. 
Together, these two forms of parallelism constitute a tightly-coupled parallel topology. Within this topology, all participating GPUs must operate collaboratively at a predefined, fixed scale.

\subsection{Communication and Migration Costs}
Fault recovery strategies generally fall into two categories: restart-and-reload, which incurs a minute-scale RTO due to I/O, and in-place recovery, which converts the RTO into an inter-device communication problem.

After a failure, the core of system recovery is the reorganization of surviving GPU resources~\cite{blagoev2025all}. This process inevitably involves data migration, including model parameters and the KV cache. The efficiency of this migration directly determines recovery time, with communication bandwidth as the main bottleneck. In a typical multi-GPU server, inter-device bandwidth varies significantly: Peer-to-Peer (P2P) direct access over high-speed interconnects (IF/XGMI) is much faster than host-mediated PCIe paths relayed through CPU memory~\cite{nakamura2020beneficial}. Therefore, an efficient plan must be topology-aware and maximize high-bandwidth paths to minimize recovery costs.

\subsection{Recovery objectives and practical metrics}
For online fault tolerance, we ground recovery objectives in measurable quantities. We use Time to First Success (TFS) as the operational form of the recovery objective: the time from fault injection to the first successful response. We report Time to Peak (TTP) as the time from TFS until throughput stabilizes at its peak. Consistency and freshness are reflected by state-retention coverage and reload fraction rather than a standalone RPO metric. In particular, we report (i) data reuse ratio and reload fraction for parameters and KVs, and (ii) the P2P/Host/Reload migration breakdown and overlap ratio during execution.

\section{Methodology}
\label{sec:design}

\begin{figure}[t]
  \centering
  \includegraphics[width=\linewidth]{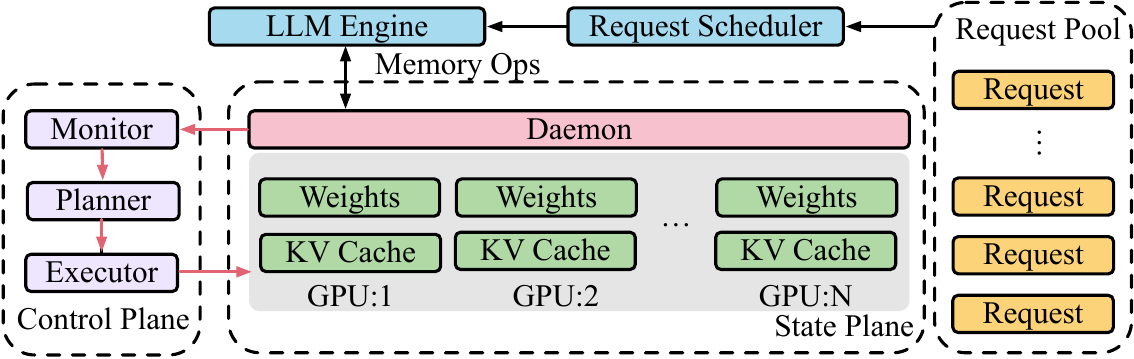}
  \vspace{-1.5em}
  \caption{AnchorTP overview with two planes. The state plane runs daemons that pin GPU memory for model parameters and the KV cache. The control plane monitors failures, plans recovery via our Continuous Minimal Migration (CMM) algorithm, and the executor coordinates data migration and system reinitialization.}
  \label{fig:anytp}
  \vspace{-1.5em}
\end{figure}

We propose AnchorTP, an architecture for state-preserving recovery and elastic tensor parallelism shown in \Cref{fig:anytp}. Its core idea is to decouple long-lived state management from dynamic topology orchestration with two planes: a state plane that pins parameters/KV, and a control plane that plans with CMM and executes recovery.

Under normal operation, daemons in the state plane hold references to the model weights and the key-value cache (KV cache). If a GPU fails, the inference process may crash; however, because the daemons persist, the GPU driver does not reclaim this critical state. The control plane then activates: it detects the failure through heartbeat checks and performance monitoring, assesses surviving GPU resources, and leverages elastic tensor parallelism to derive a viable new topology. Using our continuous minimal migration algorithm, it computes an optimal data-migration plan that maximizes data reuse while minimizing host reloads. The control plane then coordinates the actual migration, prioritizing P2P transfers and pre-allocating destination buffers to minimize recovery time. Finally, it launches a new inference system instance to resume service.

\subsection{State Management}

Traditional restart-and-reload is the most straightforward response to failures, but it yields a long TFS. We decompose it into multiple stages: init, load, and warmup. Among these, loading model state from disk into GPU memory is the primary bottleneck. For a 30B model with MoE in half precision roughly needs 60 GB of weights and typically takes tens of seconds even on fast hardware. Eliminating disk reload is therefore key to achieving second-level TFS.

To address this, we decouple a long-lived daemon from the inference system. Our design leverages the common observation that production LLM services often have significant memory redundancy, as VRAM is provisioned for KV cache usage. In our prototype, we provide a state plane and a device-memory pool interface that hold critical GPU memory regions. The daemon's role is simple and stable: at service startup it allocates or registers the required GPU memory and remains the owner of these allocations. This pinning of state has minimal overhead as it occupies otherwise under-utilized VRAM. The daemon itself does not participate in complex computation or communication, maximizing its stability.

The inference system acts as the user of these memories. Via inter-process communication (IPC), it acquires handles from the daemon and performs reads and writes.

Furthermore, the state plane maintains backup timestamps and recent usage information for each session's KV cache. During the recovery phase, if the calculated KV demand exceeds the memory budget, the system will gradually release cold KVs using a least recently used (LRU) eviction policy to prioritize the availability of hot sessions and new requests. When necessary, only the minimum set of hotspots that can cover steady-state throughput will be retained, and on-demand replay will be triggered for evicted sessions to rebuild the KV.

As a result, model parameters and KV-cache state on surviving GPUs are preserved in place. When the control-plane orchestrator launches new inference processes, they can immediately obtain handles to the surviving memory from the daemon and execute the recovery plan.

\subsection{Elastic Tensor Parallelism}

\begin{algorithm}[t]
  \caption{Continuous Minimal Migration Algorithm}
  \label{alg:cmm}
  \begin{algorithmic}[1]
  \Require 
  \Statex Current row plan: $\{(gpu_i, s_i, e_i, alive_i)\}_{i=1}^{N}$ from daemon,
  \Statex target GPU count $M$
  \Ensure Migration plan: $\{(gpu_j, sources_j)\}_{j=1}^{M}$
  \State $H \gets \max(e_i)$ for all $i$;
  \For{$j = 1$ to $M$}
    \State $s_j \gets \lfloor (j-1) \times H / M \rfloor$;
    \State $e_j \gets \lfloor j \times H / M \rfloor$;
  \EndFor
  \For{$j = 1$ to $M$}
    \State $tar\_range \gets [s_j, e_j)$;
    \For{$i = 1$ to $N$}
      \If{$alive_i = \texttt{true}$}
                 \State $inter \gets tar\_range \cap [s_i, e_i)$;
         \If{$inter \neq \emptyset$}
           \State $\texttt{AddTransferPlan}(inter, gpu_i, gpu_j)$;
           \State $tar\_range \gets tar\_range \setminus inter$;
         \EndIf
       \EndIf
     \EndFor
     \If{$tar\_range \neq \emptyset$}
       \State $\texttt{AddReloadPlan}(tar\_range, gpu_j)$;
     \EndIf
  \EndFor
  \State \Return migration plan;
  \end{algorithmic}
\end{algorithm}


Elastic tensor parallelism is a core capability that enables AnchorTP to achieve fault recovery. Traditional tensor parallelism assumes the tensor dimension is divisible by the parallel degree. Our framework removes this constraint by allowing partitioning across any number of GPUs, even if shards have unequal sizes.

Concretely, during partitioning we allow some GPUs to hold shard size \(\lfloor S / g_t \rfloor\) while others hold \(\lceil S / g_t \rceil\), where \(S\) is the tensor dimension and \(g_t\) is the new TP size. This flexibility underpins elastic recovery: regardless of how many GPUs survive after a failure, the system can always find a valid sharding plan without being constrained by divisibility.

To make the above partitioning reusable during recovery, we establish several layout invariants on the state plane: parameters reside as contiguous blocks in a daemon-managed device-memory pool and are exposed via versioned handles that provide a unified address view; the KV cache is localized by attention head-groups per token block and kept consistent with the parallel mapping. On top of these invariants, we provide two minimal interface primitives: a parameter re-assignment/reloading primitive, and KV-cache operators that support variable-length shards. Both primitives are decoupled from communication and memory management, enabling in-place remapping within a freeze window so that mapping switches complete with minimal migration.

\subsection{Topology-aware recovery planning}

\begin{figure}[t]
  \centering
    \vspace{-0.5em}
  \includegraphics[width=0.9\linewidth]{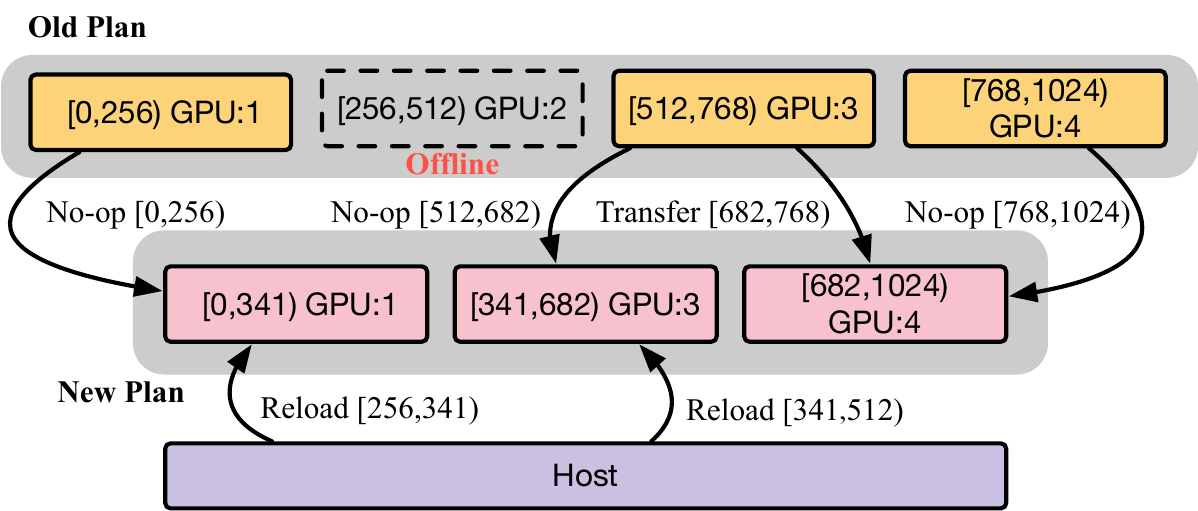}
    \vspace{-0.5em}
    \caption{Example (4$\rightarrow$3 GPUs). 1024 rows (modeled as a 1D byte interval) are split across 4 GPUs. After GPU:2 fails, the target plan is $[0,341)$, $[341,682)$, $[682,1024)$. GPU:1 keeps $[0,256)$ and reloads $[256,341)$; GPU:3 reloads $[341,512)$ and keeps $[512,682)$; GPU:4 receives $[682,768)$ via P2P from GPU:3 and keeps $[768,1024)$. Only 256 rows are reloaded; the rest use P2P.}
  \label{fig:reshard}
    \vspace{-1.5em}
  \end{figure}

Building on state preservation, we design a pragmatic two-stage recovery procedure. The first stage, logical migration planning, is topology-agnostic. It operates on a unified one-dimensional byte layout and emits a minimal-reload plan by maximally reusing surviving data intervals. The second stage, physical execution scheduling, is topology-aware. It takes the logical plan and orchestrates data movement to minimize wall-clock recovery time, respecting hardware constraints like peer reachability and bandwidth hierarchy. This decoupled design keeps the planning algorithm simple and provably optimal in terms of reload bytes, while the scheduler handles the complexities of hardware performance.

As illustrated in \Cref{fig:reshard}, the CMM algorithm first constructs the old layout view in a unified one-dimensional byte space. It identifies the total size $H = \max(e_i)$ and evenly partitions the interval $[0, H)$ by the target GPU count $M$ to derive the target layout. Then, for each target interval, it computes intersections with surviving source intervals. Non-empty intersections are mapped to P2P Transfer tasks. Any remaining gaps in the target interval that are not covered by surviving data are mapped to Reload tasks from host memory. This greedy, interval-based procedure is simple, deterministic, and guarantees a minimal total reload volume, as formally argued in \Cref{eq:min-reload}. The overall procedure is summarized in \Cref{alg:cmm}.

We give a brief argument that the planner achieves minimal reload. Let the global space be $[0,H)$, the surviving source layout be disjoint intervals $\{[s_i,e_i)\}_{i\in S}$, and the target layout be disjoint intervals $\{[u_j,v_j)\}_{j=1}^M$ with $\bigcup_j [u_j,v_j)=[0,H)$. For each target $[u_j,v_j)$, the maximum content reusable without reload equals the total measure of $\bigcup_{i\in S}([s_i,e_i)\cap [u_j,v_j))$. Hence the minimum reload over all plans is
\begin{equation}
\begin{aligned}
\mathrm{Reload}^*\;=\;&\sum_{j=1}^M\Big(|[u_j,v_j)|-\sum_{i\in S}|[s_i,e_i)\cap [u_j,v_j)|\Big) \\
=\;& H-\sum_{i\in S}\sum_{j=1}^M|[s_i,e_i)\cap [u_j,v_j)|.
\end{aligned}
\label{eq:min-reload}
\end{equation}
The CMM planner enumerates these intersections and assigns each overlap directly, thereby attaining the upper bound on reuse for every target GPU. Its reload volume thus equals the theoretical minimum derived in \Cref{eq:min-reload}. This optimality holds under the assumption that reload cost per byte is strictly higher than any P2P transfer cost.

Next, the execution scheduler optimizes the wall-clock time for the migration plan generated by CMM. 

Before launching any data movement, the scheduler first scans the entire migration plan and pre-allocates the required destination buffers on all target GPUs in a single pass. This shifts allocator overhead off the critical recovery path and avoids latency spikes caused by on-the-fly allocations during migration.

The scheduler then pipelines migration tasks to maximize hardware utilization, informed by the underlying topology. It maintains a cost model for different communication paths and prioritizes scheduling transfers over the highest-bandwidth available links first. Crucially, it overlaps high-latency reloads (host-to-GPU transfers via PCIe) with lower-latency P2P transfers (GPU-to-GPU copies, often via XGMI), effectively hiding the P2P communication latency under the longer reload time and thus shortening the overall recovery duration. In addition, the runtime KV cache and parameters share the same planning and task generation process. On KV cache misses, we rebuild via on-demand recomputation (token replay) without blocking the recovery pipeline.

\begin{figure}[t]
  \centering
  \includegraphics[width=\linewidth]{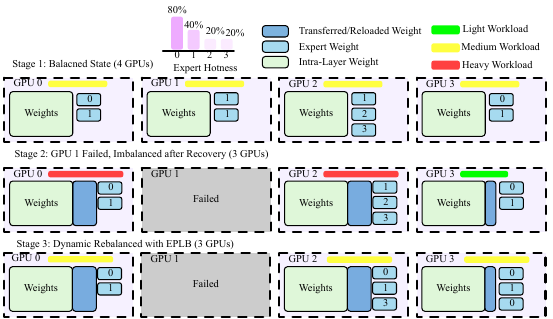}
  \vspace{-1.5em}
  \caption{Example of EPLB rebalancing after a failure. (a) Initially, 4 GPUs are perfectly balanced. (b) After GPU 1 fails, AnchorTP re-shards the parameters, leaving GPU 3 with a smaller shard and thus more free compute resources. (c) EPLB, aware of this, intelligently places the recovered Expert 1 and new replicas of the hotspot Expert 0 onto the most idle GPU (GPU 3), achieving a new, performance-optimal state that is not arithmetically balanced but maximizes system throughput.}
  \label{fig:eplb_rebalance}
  \vspace{-1.5em}
\end{figure}

Beyond failure recovery, this two-stage planning mechanism also underpins dynamic rescheduling during normal operation. When the system detects hardware performance drift, resource bottlenecks, or load hotspots, it proactively triggers rescheduling and reuses the same planning and optimization pipeline. To address potential MoE expert imbalance after elastic adjustments, the system applies an EPLB-like~\cite{guo2025deepseek} dynamic load balancing strategy. This strategy restores balance by replicating hot-spot experts and adjusting routing to optimally utilize available resources. For instance, after an $8 \to 7$ GPU recovery, the new parameter sharding might leave one GPU with a smaller shard and thus more idle compute capacity. As illustrated in \Cref{fig:eplb_rebalance}, an intelligent EPLB planner detects this heterogeneity and schedules more expert replicas to the under-utilized GPU. This results in a new state that, while not arithmetically balanced in terms of request count, is performance-optimal as it maximizes the throughput of the entire system.

\section{Implementation}
\label{sec:implementation}

We implement a lightweight inference framework based on nano-vllm~\cite{nanovllm} and realize the aforementioned fault-tolerance design in practice. On the state plane, a daemon manages a device-memory pool and IPC handles to anchor model parameters and the KV cache. The migration planner operates over a unified one-dimensional byte space and generates Transfer/Reload tasks on interval boundaries. The orchestrator distributes versioned plans, while the executor pre-allocates destination buffers and overlaps Reloads with P2P transfers. ETP in AnchorTP enables unequal-width sharding for attention and linear layers, making arbitrary TP degrees feasible and allowing reuse of surviving shards. The implementation follows minimal-interface and decoupling principles: communication, memory, and operator changes are independent; stale shards are promptly released after remapping takes effect to curb fragmentation, which eases integration into existing services. For model reload, our loader reads safetensors by offsets: it parses per-tensor offset/length from the header and issues range reads to fetch only required byte intervals.

\section{Evaluation}
\label{sec:evaluation}

This chapter systematically evaluates AnchorTP's fault-recovery capability and elastic scaling efficiency on a single-node, multi-GPU setup, covering recovery latency, resource utilization, and end-to-end performance.
We evaluate recovery in 4$\rightarrow$2 and 8$\rightarrow$6 TP-degradation scenarios, and additionally test 8$\rightarrow$7 for ablations.

\begin{table*}[t]
  \centering
  \caption{End-to-end recovery performance comparison. We report TFS and TTP (in seconds) after failures injected at 25\% and 50\% of the request stream. `Overhead` is the total runtime increase over the no-failure Static Parallelism (SP) baseline; lower is better.}
  \label{tab:e2e}
  \setlength{\tabcolsep}{8pt}
  \renewcommand{\arraystretch}{1.6}
  \resizebox{1\textwidth}{!}{
  \begin{tabular}{@{}lllrrrrr|r}
    \toprule
    Model & TP $\Delta$ & Method & TFS@25 (s) & TFS@50 (s) & TTP@25 (s) & TTP@50 (s) & Total (s) & Overhead (s) $\downarrow$ \\
    \midrule
    \multirow{3}{*}{Qwen3-30B-A3B} & & \cellcolor{gray!10}Static Parallelism (SP) & \cellcolor{gray!10} / & \cellcolor{gray!10} / & \cellcolor{gray!10} / & \cellcolor{gray!10} / & \cellcolor{gray!10} 107.84$\pm$5.68 & \cellcolor{gray!10}0 (Baseline) \\
    \cline{2-9}
    & \multirow{2}{*}{4 $\rightarrow$ 2} & AnchorTP (ours) & 4.48$\pm$0.35 & 4.98$\pm$0.43 & 6.23$\pm$0.82 & 7.91$\pm$1.07 & \cellcolor{green!10}139.74$\pm$6.76 & \cellcolor{green!10}+31.90 \\
      &  & Elastic TP (restart-only) & 48.43$\pm$2.95 & 49.21$\pm$3.17 & 14.15$\pm$1.74 & 19.33$\pm$2.11 & \cellcolor{red!10}285.82$\pm$9.81 & \cellcolor{red!10}+177.98 \\
    \midrule
    \multirow{3}{*}{Mixtral-8$\times$22B} & & \cellcolor{gray!10}Static Parallelism (SP) & \cellcolor{gray!10} / & \cellcolor{gray!10} / & \cellcolor{gray!10} / & \cellcolor{gray!10} / & \cellcolor{gray!10} 273.65$\pm$8.42 & \cellcolor{gray!10}0 (Baseline) \\
    \cline{2-9}
    & \multirow{2}{*}{8 $\rightarrow$ 6} & AnchorTP (ours) & 18.71$\pm$1.44 & 19.52$\pm$1.12 & 25.33$\pm$2.34 & 28.94$\pm$2.88 & \cellcolor{green!10}345.12$\pm$9.67 & \cellcolor{green!10}+71.47 \\
      & & Elastic TP (restart-only) & 195.82$\pm$5.07 & 198.49$\pm$6.53 & 35.69$\pm$3.53 & 41.27$\pm$3.71 & \cellcolor{red!10}610.79$\pm$17.61 & \cellcolor{red!10}+337.14 \\
     \bottomrule
  \end{tabular}
    }
  \\
\end{table*}

\minisection{Experimental Setup}
The evaluation is conducted on a single-node, multi-GPU ROCm 6.3 platform to first establish the foundational performance of our recovery primitives in a controlled environment with high-speed interconnects. The setup consists of $8\times$ AMD Instinct MI210 with 64GB memory and dual NUMA (GPUs 0--3 on node0 and GPUs 4--7 on node1). Intra-group interconnect is via Infinity Fabric (IF/XGMI), inter-group connectivity is over PCIe, and RCCL is used as the communication library. The software stack is Python 3.12 and PyTorch 2.8.0.

We use the following metrics: TFS, defined as the time from fault injection to the first successful response; TTP, the time from TFS until throughput stabilizes at its peak; and the total run time of the benchmark. During the recovery window (from failure detection to TFS), incoming requests are queued by the serving frontend and processed once the service is restored, ensuring no requests are dropped.

Models and Workload. We use Qwen3-30B-A3B~\cite{qwen3technicalreport}, Mixtral-8$\times$22B (141B)~\cite{jiang2024mixtral}, and Qwen3-8B/14B~\cite{qwen3technicalreport}. The workload is a fixed set of 1,000 ShareGPT requests~\cite{sharegpt} replayed with a fixed arrival pattern.

Baselines. Static Parallelism (SP) keeps the original TP size without online rescheduling; upon failure, service halts (we report its no-failure runtime as a baseline reference). Elastic TP (restart-only) restarts with arbitrary TP sizes (supports non-divisible TP) but performs no state-preserving recovery (full reload at restart).

\minisection{End-to-End Performance}
We evaluate end-to-end performance under injected failures at 25\% and 50\% of the request stream.

For Qwen3-30B-A3B, we simulate one GPU failure at 25\% and another at 50\% of the request stream.
For Qwen3-30B-A3B, all GPUs are in the same IF group. For Mixtral-8$\times$22B (8$\rightarrow$6), failures are induced symmetrically with one GPU terminated in each NUMA node's IF group to ensure balanced degradation.

\begin{table}[t]
  \centering
  \footnotesize
  \caption{Planner comparison on Reload and P2P time (Mixtral-8$\times$22B, 8$\to$7 GPUs). Lower is better.}
  \label{tab:planner-simple}
  \renewcommand{\arraystretch}{1.1}
    {%
  \begin{tabular}{l rr}
  \toprule
  Method & Reload Time (s) $\downarrow$ & P2P Time (s) $\downarrow$ \\
  \midrule
  CM (ours)   & \cellcolor{green!10}17.56$\pm$1.39 & 1.87$\pm$0.33 \\
  Greedy      & \cellcolor{red!10}26.15$\pm$1.73 & / \\
  Full Reload & \cellcolor{red!25}197.31$\pm$7.98 & / \\
  \bottomrule
  \end{tabular}%
  }
  \vspace{-0.25cm}
\end{table}

\minisection{Ablation Study}
\label{sec:ablation_study}
SP cannot recover from failure; `/' indicates TFS/TTP are undefined.
Even when KV-cache misses occur and trigger on-demand replay, AnchorTP degrades gracefully and continues serving, which is strictly preferable to service interruption in SP.
TFS of AnchorTP is about $10.8\times$ (25\%) and $9.9\times$ (50\%) faster for Qwen3-30B-A3B, and approximately $10.5\times$ faster for Mixtral-8$\times$22B, compared to the Elastic TP baseline, as it significantly reduces the costly full model reload from host memory.
This efficiency translates directly to a significantly lower total runtime overhead, which is reduced by factors of $5.6\times$ and $4.7\times$ for Qwen3-30B-A3B and Mixtral-8$\times$22B, respectively.
As our analysis, the TFS for the restart-only baseline is dictated by the total model size, with Mixtral-8$\times$22B taking roughly four times longer to load than Qwen3-30B-A3B, consistent with its larger parameter count. In contrast, AnchorTP's TFS is determined by minimal data migration, resulting in consistently low recovery times.

\begin{figure}
  \centering
  \vspace{0.3cm}
  \includegraphics[width=0.75\linewidth]{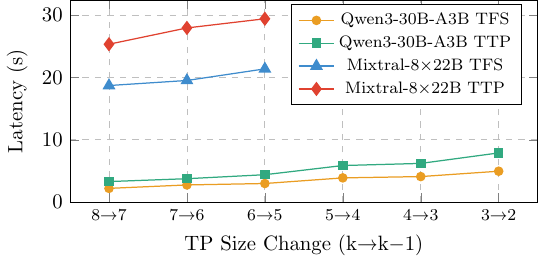}
  \caption{Per-switch TFS and TTP as TP decreases (k$\rightarrow$k$-$1) for Qwen3-30B-A3B and Mixtral-8$\times$22B. Lower is better.}
  \label{fig:scale_generalization}
  \vspace{-0.25cm}
\end{figure}

To independently validate the contributions of AnchorTP's core components, a series of ablation studies were conducted to isolate the impact of each component. First, the effectiveness of daemon-based state management is assessed by comparing the full system against the "Elastic TP (restart-only)" baseline. As shown in \Cref{tab:e2e}, removing state preservation forces a full reload, significantly increasing TFS and TTP and underscoring the daemon's pivotal role. Second, the "Static Parallelism (SP)" baseline in \Cref{tab:e2e} illustrates the consequence of lacking elastic reconfiguration: upon failure, the service cannot adapt and is irrecoverably interrupted, establishing elastic TP as a foundational prerequisite for fault tolerance.

The superiority of our Continuous Minimal Migration algorithm is evaluated against alternative strategies. The results in \Cref{tab:planner-simple} demonstrate its efficiency in minimizing reload and P2P transfer times. The evaluated methods include CM (ours), a Greedy algorithm with local optimizations, and a Full Reload approach without data reuse. For Greedy in our experiments, missing bytes are reloaded from host memory on the target GPU and we do not schedule cross-device transfers for those gaps. In numbers, CM achieves the lowest reload time $\approx17.6$s while incurring only small P2P ($\approx1.9$s); Greedy increases reload time to $\approx26.2$s because gaps are not filled via P2P; Full Reload is dominated by host reload ($\approx197$s) and has no P2P component, corroborating the advantage of reuse-aware planning.

To assess AnchorTP's performance stability under continuous degradation and its adaptability to different model scales, we progressively decrease the TP degree for two models: Qwen3-30B-A3B (from TP=8 to 2) and Mixtral-8$\times$22B (from TP=8 to 6). For each change (TP=k $\rightarrow$ k$-$1), we report per-switch TFS and TTP, as shown in \Cref{fig:scale_generalization}.

As depicted in \Cref{fig:scale_generalization}, both TFS and TTP exhibit a consistent upward trend as the Tensor Parallelism (TP) degree is progressively reduced. This trend is attributed to the increased complexity of rescheduling and the higher load placed on each surviving GPU. Notably, TFS shows a moderate, near-linear increase, demonstrating the efficiency of our minimal migration algorithm in handling state reconstruction across various scales. In contrast, TTP rises more steeply, underscoring the inherent performance challenge as fewer devices must handle the same workload, prolonging the time required to stabilize at a new peak throughput. The significantly higher absolute latencies for Mixtral-8$\times$22B compared to Qwen3-30B-A3B further highlight the pronounced impact of model scale on recovery overhead. Overall, the results validate AnchorTP's robust scalability, maintaining efficient recovery even under significant degradation.

\begin{figure}
    \centering
    \includegraphics[width=0.8\linewidth]{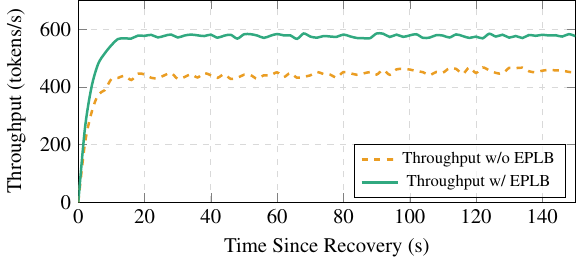}
    \caption{Impact of EPLB on system throughput. With EPLB enabled, the system not only reaches a higher peak throughput but also stabilizes much faster, as indicated by the shorter TTP window. This demonstrates that rebalancing accelerates the convergence to a new, optimal steady-state.}
    \label{fig:moe_throughput}
    \vspace{-1.5em}
\end{figure}

When serving MoE models like Mixtral-8$\times$22B under full load, elastic recovery ($8 \to 7$ GPUs) induces load imbalance from stale expert mappings. A lightweight EPLB strategy~\cite{guo2025deepseek} restores balance: without EPLB the system stabilizes at 436.61 tokens/s, while with EPLB it reaches 562.32 tokens/s (+29\%) and shortens TTP (\Cref{fig:moe_throughput}).

\section{Related Work}

\minisection{LLM Distributed Systems}
LLM training and inference rely on advanced parallelism including Data Parallelism (DP), Tensor Parallelism (TP), Pipeline Parallelism (PP), and Fully Sharded Data Parallel (FSDP).
Inference stacks such as vLLM, TensorRT-LLM, and Megatron variants provide high-throughput kernels and KV management.
DreamDDP~\cite{tang2025dreamddpacceleratingdataparallel} accelerates DP under low-bandwidth settings by partially localizing SGD and overlapping synchronization;
AsymGroup~\cite{AsymGroup} dynamically forms asymmetric 2D DP groups to manage heterogeneity.

\minisection{Online Disaster Recovery and Elasticity for Serving}
As LLM serving scales, failures become common and elasticity is required.
Adaptive fault tolerance leverages prediction and dynamic resource allocation~\cite{jin2025adaptivefaulttolerancemechanisms}.
For training, Nonuniform Tensor Parallelism (NTP) mitigates single-GPU failures by reconfiguring TP within a DP replica via gradient resharding~\cite{arfeen2025nonuniformtensorparallelismmitigatinggpufailure}.
Resource managers like FaPES~\cite{FaPES} borrow GPUs across inference and training to improve utilization. Unlike these, AnchorTP targets inference-time recovery under TP-scale changes by preserving state in-place and performing minimal migration.

\minisection{Parameter/KV Hot Loading for Serving}
Full-model reload dominates switch latency in large models.
Parameter hot loading decouples model architecture from parameters to accelerate switching (FastPTM~\cite{FastPTM}).
KV-centric serving further reduces memory churn in vLLM-like systems.
AnchorTP complements these by pinning parameters and KVs in GPU memory via a daemon and by enabling elastic TP so that surviving shards are reused rather than reloaded.

\section{Discussion}

\minisection{Assumptions and Limitations}
Our design targets single-node, multi-GPU inference with peer reachability within the node. The reload-minimality of CMM assumes per-byte costs where reload is strictly more expensive than transfer/no-op and ignores link reachability during planning; execution scheduling then respects hardware topology. Extending CMM to multi-node settings requires incorporating link reachability and cost weighting into planning.

\minisection{Load Rebalancing and MoE Dynamics}
Post-recovery, topology changes can perturb expert load. We integrate EPLB to re-stabilize routing weights with lightweight regulation, prioritizing intra-node balance before inter-GPU adjustments. This complements CMM by restoring steady-state performance without invalidating preserved KVs.

\section{Conclusion}
\label{sec:conclusion}

This paper introduced AnchorTP, a disaster recovery framework designed to address the high-availability challenges in Tensor Parallelism-based LLM inference services. Traditional approaches suffer from service interruptions due to fixed device scales and long downtimes from state reloading. AnchorTP overcomes these limitations through two key innovations: Elastic Tensor Parallelism, which enables flexible service reconfiguration on surviving GPUs, and State-Preserving Recovery, which uses a decoupled daemon to preserve critical states in GPU memory, thus eliminating costly reloads. A topology-aware migration planner further accelerates recovery by minimizing data movement.

Experiments demonstrate that AnchorTP is highly effective, reducing recovery time by over 10$\times$ and time-to-peak-performance by up to 59\% compared to restart methods. This work significantly enhances the resilience of LLM inference services. Future work will extend CMM to multi-node topologies with link reachability and cost weighting, and integrate predictive fault handling for preemptive warm migration.

\clearpage
\bibliographystyle{IEEEtran}
\bibliography{ref/refs}

@misc{tang2025dreamddpacceleratingdataparallel,
      title={DreamDDP: Accelerating Data Parallel Distributed LLM Training with Layer-wise Scheduled Partial Synchronization}, 
      author={Zhenheng Tang and Zichen Tang and Junlin Huang and Xinglin Pan and Rudan Yan and Yuxin Wang and Amelie Chi Zhou and Shaohuai Shi and Xiaowen Chu and Bo Li},
      year={2025},
      eprint={2502.11058},
      archivePrefix={arXiv},
      primaryClass={cs.DC},
      url={https://arxiv.org/abs/2502.11058}, 
}

@ARTICLE{AsymGroup,
  author={Tae Kim, Ki and Im, Seok-Ju and Chung, Eui-Young},
  journal={IEEE Access}, 
  title={AsymGroup: Asymmetric Grouping and Communication Optimization for 2D Tensor Parallelism in LLM Inference}, 
  year={2025},
  volume={13},
  number={},
  pages={120591-120602},
  doi={10.1109/ACCESS.2025.3587387}}

@misc{jin2025adaptivefaulttolerancemechanisms,
      title={Adaptive Fault Tolerance Mechanisms of Large Language Models in Cloud Computing Environments}, 
      author={Yihong Jin and Ze Yang and Xinhe Xu and Yihan Zhang and Shuyang Ji},
      year={2025},
      eprint={2503.12228},
      archivePrefix={arXiv},
      primaryClass={cs.DC},
      url={https://arxiv.org/abs/2503.12228}, 
}

@misc{arfeen2025nonuniformtensorparallelismmitigatinggpufailure,
      title={Nonuniform-Tensor-Parallelism: Mitigating GPU failure impact for Scaled-up LLM Training}, 
      author={Daiyaan Arfeen and Dheevatsa Mudigere and Ankit More and Bhargava Gopireddy and Ahmet Inci and Gregory R. Ganger},
      year={2025},
      eprint={2504.06095},
      archivePrefix={arXiv},
      primaryClass={cs.DC},
      url={https://arxiv.org/abs/2504.06095}, 
}

@inproceedings{FaPES,
author = {Zhao, Xiaoyang and Yang, Siran and Wang, Jiamang and Diao, Lansong and Qu, Lin and Wu, Chuan},
title = {FaPES: Enabling Efficient Elastic Scaling for Serverless Machine Learning Platforms},
year = {2024},
isbn = {9798400712869},
publisher = {Association for Computing Machinery},
address = {New York, NY, USA},
url = {https://doi.org/10.1145/3698038.3698548},
doi = {10.1145/3698038.3698548},
booktitle = {Proceedings of the 2024 ACM Symposium on Cloud Computing},
pages = {443–459},
numpages = {17},
location = {Redmond, WA, USA},
series = {SoCC '24}
}

@article{FastPTM,
title = {FastPTM: Fast weights loading of pre-trained models for parallel inference service provisioning},
journal = {Parallel Computing},
volume = {122},
pages = {103114},
year = {2024},
issn = {0167-8191},
doi = {https://doi.org/10.1016/j.parco.2024.103114},
url = {https://www.sciencedirect.com/science/article/pii/S0167819124000528},
author = {Fenglong Cai and Dong Yuan and Zhe Yang and Yonghui Xu and Wei He and Wei Guo and Lizhen Cui}
}

@inproceedings{wu2024loongserve,
  title={Loongserve: Efficiently serving long-context large language models with elastic sequence parallelism},
  author={Wu, Bingyang and Liu, Shengyu and Zhong, Yinmin and Sun, Peng and Liu, Xuanzhe and Jin, Xin},
  booktitle={Proceedings of the ACM SIGOPS 30th Symposium on Operating Systems Principles},
  pages={640--654},
  year={2024}
}

@inproceedings{wang2022tesseract,
  title={Tesseract: Parallelize the tensor parallelism efficiently},
  author={Wang, Boxiang and Xu, Qifan and Bian, Zhengda and You, Yang},
  booktitle={Proceedings of the 51st International Conference on Parallel Processing},
  pages={1--11},
  year={2022}
}

@article{brakel2024model,
  title={Model parallelism on distributed infrastructure: A literature review from theory to LLM case-studies},
  author={Brakel, Felix and Odyurt, Uraz and Varbanescu, Ana-Lucia},
  journal={arXiv preprint arXiv:2403.03699},
  year={2024}
}

@article{arfeen2025nonuniform,
  title={Nonuniform-Tensor-Parallelism: Mitigating GPU failure impact for Scaled-up LLM Training},
  author={Arfeen, Daiyaan and Mudigere, Dheevatsa and More, Ankit and Gopireddy, Bhargava and Inci, Ahmet and Ganger, Gregory R},
  journal={arXiv preprint arXiv:2504.06095},
  year={2025}
}

@article{cui2025characterizing,
  title={Characterizing gpu resilience and impact on ai/hpc systems},
  author={Cui, Shengkun and Patke, Archit and Nguyen, Hung and Ranjan, Aditya and Chen, Ziheng and Cao, Phuong and Bode, Brett and Bauer, Gregory and Di Martino, Catello and Jha, Saurabh and others},
  journal={arXiv preprint arXiv:2503.11901},
  year={2025}
}

@article{blagoev2025all,
  title={All is Not Lost: LLM Recovery without Checkpoints},
  author={Blagoev, Nikolay and Ersoy, O{\u{g}}uzhan and Chen, Lydia Yiyu},
  journal={arXiv preprint arXiv:2506.15461},
  year={2025}
}

@inproceedings{xia2025mnemosyne,
  title={Mnemosyne: Lightweight and Fast Error Recovery for LLM Training in a Just-In-Time Manner},
  author={Xia, Jinyi and Zhang, Menghao and Huang, Jiaxun and Liu, Yuezheng and Hu, Xiaohe and Liu, Xudong and Hu, Chunming},
  booktitle={Proceedings of the 9th Asia-Pacific Workshop on Networking},
  pages={157--163},
  year={2025}
}

@article{zheng2024sglang,
  title={Sglang: Efficient execution of structured language model programs},
  author={Zheng, Lianmin and Yin, Liangsheng and Xie, Zhiqiang and Sun, Chuyue Livia and Huang, Jeff and Yu, Cody Hao and Cao, Shiyi and Kozyrakis, Christos and Stoica, Ion and Gonzalez, Joseph E and others},
  journal={Advances in neural information processing systems},
  volume={37},
  pages={62557--62583},
  year={2024}
}

@inproceedings{kwon2023efficient,
  title={Efficient memory management for large language model serving with pagedattention},
  author={Kwon, Woosuk and Li, Zhuohan and Zhuang, Siyuan and Sheng, Ying and Zheng, Lianmin and Yu, Cody Hao and Gonzalez, Joseph and Zhang, Hao and Stoica, Ion},
  booktitle={Proceedings of the 29th symposium on operating systems principles},
  pages={611--626},
  year={2023}
}

@article{strati2024d,
  title={D$\backslash$'ej$\backslash$avu: Kv-cache streaming for fast, fault-tolerant generative llm serving},
  author={Strati, Foteini and Mcallister, Sara and Phanishayee, Amar and Tarnawski, Jakub and Klimovic, Ana},
  journal={arXiv preprint arXiv:2403.01876},
  year={2024}
}

@article{wu2023transom,
  title={Transom: An efficient fault-tolerant system for training llms},
  author={Wu, Baodong and Xia, Lei and Li, Qingping and Li, Kangyu and Chen, Xu and Guo, Yongqiang and Xiang, Tieyao and Chen, Yuheng and Li, Shigang},
  journal={arXiv preprint arXiv:2310.10046},
  year={2023}
}

@article{xie2025realm,
  title={ReaLM: Reliable and Efficient Large Language Model Inference with Statistical Algorithm-Based Fault Tolerance},
  author={Xie, Tong and Zhao, Jiawang and Wan, Zishen and Zhang, Zuodong and Wang, Yuan and Wang, Runsheng and Huang, Ru and Li, Meng},
  journal={arXiv preprint arXiv:2503.24053},
  year={2025}
}

@article{cheng2025scalable,
  title={A Scalable Approach to Distributed Large Language Model Inference},
  author={Cheng, Yihua},
  year={2025},
  publisher={University of Chicago},
  journal={knowledge.uchicago.edu}
}

@article{xu2025cloud,
  title={Cloud Native System for LLM Inference Serving},
  author={Xu, Minxian and Liao, Junhan and Wu, Jingfeng and He, Yiyuan and Ye, Kejiang and Xu, Chengzhong},
  journal={arXiv preprint arXiv:2507.18007},
  year={2025}
}

@article{zhang2024edgeshard,
  title={Edgeshard: Efficient llm inference via collaborative edge computing},
  author={Zhang, Mingjin and Shen, Xiaoming and Cao, Jiannong and Cui, Zeyang and Jiang, Shan},
  journal={IEEE Internet of Things Journal},
  year={2024},
  publisher={IEEE}
}

@article{jacobs1991adaptive,
  title={Adaptive mixtures of local experts},
  author={Jacobs, Robert A and Jordan, Michael I and Nowlan, Steven J and Hinton, Geoffrey E},
  journal={Neural computation},
  volume={3},
  number={1},
  pages={79--87},
  year={1991},
  publisher={MIT Press}
}

@article{masoudnia2014mixture,
  title={Mixture of experts: a literature survey},
  author={Masoudnia, Saeed and Ebrahimpour, Reza},
  journal={Artificial Intelligence Review},
  volume={42},
  number={2},
  pages={275--293},
  year={2014},
  publisher={Springer}
}

@article{liu2024deepseek,
  title={Deepseek-v3 technical report},
  author={Liu, Aixin and Feng, Bei and Xue, Bing and Wang, Bingxuan and Wu, Bochao and Lu, Chengda and Zhao, Chenggang and Deng, Chengqi and Zhang, Chenyu and Ruan, Chong and others},
  journal={arXiv preprint arXiv:2412.19437},
  year={2024}
}

@article{zeng2025efficientmoe,
  title={EfficientMoE: Optimizing Mixture-of-Experts Model Training With Adaptive Load Balance},
  author={Zeng, Yan and Huang, Chengchuang and Mei, Yipeng and Zhang, Lifu and Su, Teng and Ye, Wei and Shi, Wenqi and Wang, Shengnan},
  journal={IEEE Transactions on Parallel and Distributed Systems},
  year={2025},
  publisher={IEEE}
}

@phdthesis{wisdom2024load,
  title={Load balancing and memory optimizations for expert parallel training of large language models},
  author={Wisdom, Daniel},
  year={2024},
  school={Massachusetts Institute of Technology}
}

@article{guo2025deepseek,
  title={Deepseek-r1: Incentivizing reasoning capability in llms via reinforcement learning},
  author={Guo, Daya and Yang, Dejian and Zhang, Haowei and Song, Junxiao and Zhang, Ruoyu and Xu, Runxin and Zhu, Qihao and Ma, Shirong and Wang, Peiyi and Bi, Xiao and others},
  journal={arXiv preprint arXiv:2501.12948},
  year={2025}
}

@inproceedings{singh2023hybrid,
  title={A hybrid tensor-expert-data parallelism approach to optimize mixture-of-experts training},
  author={Singh, Siddharth and Ruwase, Olatunji and Awan, Ammar Ahmad and Rajbhandari, Samyam and He, Yuxiong and Bhatele, Abhinav},
  booktitle={Proceedings of the 37th International Conference on Supercomputing},
  pages={203--214},
  year={2023}
}

@inproceedings{zhu2025megascale,
  title={MegaScale-Infer: Efficient Mixture-of-Experts Model Serving with Disaggregated Expert Parallelism},
  author={Zhu, Ruidong and Jiang, Ziheng and Jin, Chao and Wu, Peng and Stuardo, Cesar A and Wang, Dongyang and Zhang, Xinlei and Zhou, Huaping and Wei, Haoran and Cheng, Yang and others},
  booktitle={Proceedings of the ACM SIGCOMM 2025 Conference},
  pages={592--608},
  year={2025}
}

@inproceedings{nakamura2020beneficial,
  title={How beneficial is peer-to-peer DMA?},
  author={Nakamura, Ryo and Kuga, Yohei and Akashi, Kunio},
  booktitle={Proceedings of the 11th ACM SIGOPS Asia-Pacific Workshop on Systems},
  pages={25--32},
  year={2020}
}

@software{nanovllm,
  author = {Xingkai Yu},
  title = {nano-vllm},
  url = {https://github.com/GeeeekExplorer/nano-vllm},
  year = {2025}
}

@misc{qwen3technicalreport,
      title={Qwen3 Technical Report}, 
      author={Qwen Team},
      year={2025},
      eprint={2505.09388},
      archivePrefix={arXiv},
      primaryClass={cs.CL},
      url={https://arxiv.org/abs/2505.09388}, 
}

@misc{sharegpt,
  author = {ShareGPT},
  title = {ShareGPT},
  howpublished = {\url{https://sharegpt.com/}},
  year = {2023}
}

@article{jiang2024mixtral,
  title={Mixtral of experts},
  author={Jiang, Albert Q and Sablayrolles, Alexandre and Roux, Antoine and Mensch, Arthur and Savary, Blanche and Bamford, Chris and Chaplot, Devendra Singh and Casas, Diego de las and Hanna, Emma Bou and Bressand, Florian and others},
  journal={arXiv preprint arXiv:2401.04088},
  year={2024}
}

\end{document}